# Scale-free structural organization of oxygen interstitials in $La_2CuO_{4+y}$

Michela Fratini[1,*], Nicola Poccia[1], Alessandro Ricci[1], Gaetano Campi[1,2], Manfred Burghammer[3], Gabriel Aeppli[4] & Antonio Bianconi[1]

[1] Department of Physics, Sapienza University of Rome, Piazzale Aldo Moro 2, 00185 Roma, Italy.
[2] Institute of Crystallography, CNR, Via Salaria Km 29.300, Monterotondo Stazione, Roma, I-00016, Italy.
[3] European Synchrotron Radiation Facility, B.P. 220, F-38043 Grenoble Cedex, France.
[4] London Centre for Nanotechnology and Department of Physics and Astronomy, University College London, 17–19 Gordon Street, London WC1H 0AH, UK.
* Present address: Institute for Photonic and Nanotechnologies, CNR, Via Cineto Romano 42,

**It is well known that the microstructures of the transition-metal oxides[1–3], including the high-transition-temperature (*high-$T_c$*) copper oxide superconductors[4–7], are complex. This is particularly so when there are oxygen interstitials or vacancies[8], which influence the bulk properties. For example, the oxygen interstitials in the spacer layers separating the superconducting $CuO_2$ planes undergo ordering phenomena in $Sr_2O_{1+y}CuO_2$ (ref. 9), $YBa_2Cu_3O_{6+y}$ (ref. 10) and $La_2CuO_{4+y}$ (refs 11–15) that induce enhancements in the transition temperatures with no changes in hole concentrations. It is also known that complex systems often have a scale-invariant structural organization[16], but hitherto none had been found in high-$T_c$ materials. Here we report that the ordering of oxygen interstitials in the $La_2O_{2+y}$ spacer layers of $La_2CuO_{4+y}$ high-$T_c$ superconductors is characterized by a fractal distribution up to a maximum limiting size of 400 mm. Intriguingly, these fractal distributions of dopants seem to enhance superconductivity at high temperature.**

To make progress in the engineering and science of multiscale phase separation, visualization is essential. Momentum-space probes such as X-ray diffraction (XRD), angle-resolved photoelectron spectroscopy and inelastic neutron scattering are excellent for characterizing order and coherent excitations but are highly ambiguous when different phases coexist. Therefore, to image micrometre-scale lattice fluctuations in high-temperature superconductors due to the ordering of oxygen interstitials (i-O), we have used scanning synchrotron radiation X-ray microdiffraction (μXRD)[17–19], which combines high wavenumber resolution with micrometre-scale spatial resolution, made possible by twin revolutions in X-ray optics and electron-accelerator-based X-ray sources. The technique examines the same bulk, showing interesting material functionalities such as superconductivity, but also gives a wavelet[20] mixed real- and reciprocal-space representation of the sample, where the nanostructure associated with short- and medium-range oxygen order is captured by means of diffraction and the microstructure is recorded in real space.

Optimally doped $La_2CuO_{4+y}$ is ideal for the investigation of intrinsic multiscale heterogeneity in copper oxides because the i-O are mobile[8] in the $La_2O_{2+y}$ layers intercalated between the super- conducting $CuO_2$ planes, forming a superlattice with the largest misfit strain in the copper oxides[21,22]. Therefore, the order parameters associated with mobile i-O in the spacer layers maintaining charge neutrality are expected to track the microscale phase separation in the $CuO_2$ planes. In the optimum doping range, $0.1 < y < 0.12$, a single $T_c = 40$ K superconducting phase appears but many experiments[11–15,23,24] indicate a complex magnetic, electronic and structural phase separation. Annealing the sample at 370 K, where i-O does not escape from the sample, followed by quenching below 200 K, yields a mixed state, displaying two critical temperatures [11,12], $32 < T_{c1} < 36$ K and $T_{c2} = 16$ K; we therefore call this state the $T_c = 16+32$ K phase. The mixed state has been thought to originate from the coexistence of regions with the same average doping level but with two different orderings of the dopants, with some unknown i-O self-organization; also, no information is available on the relation between this unknown superstructure and the high-temperature superconducting phase.

Our single-crystal samples of $La_2CuO_{4+y}$ are cleaved with their surfaces parallel to the $CuO_2$ planes and then mounted on the x–y translator of the μXRD instrument (Fig. 1a). The main XRD reflections of the single crystal (orthorhombic Fmmm space group) show no splitting or spatial spot-to-spot change. The reciprocal-space image recorded by the CCD on the right in Fig. 1a reveals the known satellite peaks, associated with the three-dimensional Q2 superstructure of the ordered i-O dopants[14,15], displaced from the main crystal reflections by the wavevector **q**2 with components $\Delta l = (0.497 \pm 0.005)c^*$, $\Delta k = (0.247 \pm 0.003)b^*$ and $\Delta h = (0.087 \pm 0.004)a^*$, where $a^*$, $b^*$ and $c^*$ are the reciprocal-lattice units; there is also a strong second harmonic displaced by 2**q**2 .

Mapping the x–y position dependence of the integrated satellite peak intensity (Fig. 1b) provides our key experimental discovery, namely that on the micrometre scale probed, the i-O order with nanometre-scale wavelength in real space (Fig. 1c), which is responsible for the satellite peaks, is highly inhomogeneous, even for an optimal ($T_c = 40$ K) superconducting sample of $La_2CuO_{4.1}$ .

The micrometre-scale inhomogeneity depends on the sample preparation history, which also controls the superconducting properties. In fact, the x–y position dependence of the Q2 superstructure intensity for two typical samples with $T_c = 40$ K (Fig. 2a) and $T_c = 16+32$ K (Fig. 2b), respectively, is quite different. We quantify the difference by looking at the distribution of intensities as well as the distance-dependent intensity correlations. The probability distribution (Fig. 2c) of XRD intensities shows a fat tail above the average Q2 intensity, <I>, extending to intensities 20 times larger than the average intensity. In this range, the data show a clear power-law distribution according to standard statistical physics criteria[25]. The difference between the probability distributions of the two sets of samples is also obvious from fitting the data using a power law $P(x) \propto x^{-\alpha} \exp(-x/x_0)$. The outcome is that the power-law exponent, α, is always indistinguishable from 2.6 and that the cut-off, $x_0$, is less than 10 for the $T_c = 16+32$ K materials and greater than 10 for the high-$T_c$ (40 K) materials.

The spatial intensity correlation function, G(r) (Fig. 2d), where $r = |\vec{R}_i - \vec{R}_j|$ is the distance between x–y positions $\vec{R}_k$ on the sample, as in Fig. 2a,b, does not exhibit exponential behaviour but follows a power law with a cut-off, as expected, near a critical point: $G(r) \propto r^{-\eta} \exp(-r/\xi)$, with $\eta$ = 0.3 ± 0.1 for all five samples. The correlation length, $\xi$, increases with increasing <I>, varying between 50 and 250 mm for the $T_c$ = 16+32 K samples and taking the respective values 400 ± 30 μm and 350 ± 30 μm for the two $T_c$ = 40 K samples. The results for both the intensity distribution and the two-point correlation function show that the unexpected fractal nature, associated with the measured power laws, of i-O ordering is robust, approaching a pure scale-free distribution in the sample with higher critical temperature. A decade of work at the Elettra storage ring in Trieste shows that thermal treatments (Fig. 3) can control the i-O self-assembly and detailed superconducting properties while maintaining constant oxygen content. In particular, the probability of ordered i-O domains can be suppressed by annealing at 350–380 K and enhanced by annealing at 200–300 K. After quenching the sample from 370 K to 300 K and keeping the temperature constant at 300 K, the Q2 XRD signal, spatially averaged over the full sample surface, increases rapidly as a function of time in the first few days, after which it increases slowly over very much longer durations. The superconducting transitions of samples with different levels of average ordered Q2 volume have been established using contactless single-coil inductance. The data clearly demonstrate that the low-$T_c$ (16 K) phase disappears with the fully disordered i-O produced after annealing above 370 K followed by rapid quenching. Modest annealing at 200–300 K induces weak Q2 XRD superstructure spots and an additional superconducting signature with $T_c$ = 16 K. On further annealing, the average intensity of the Q2 superstructure rises and the $T_c$ = 16 K superconducting phase gains prominence. After a very long annealing time, the integrated volume of the ordered i-O domains increases, the $T_c$ = 16 K phase disappears and a single superconducting phase with $T_c$ = 38–41 K finally dominates. Our work puts the copper oxides into the same category as other heterogeneous transition-metal oxides[1,2], where a key parameter is the misfit strain parameter of the superlattice[21,22], and better superconductivity is associated with critical percolation of oxygen order. Unexplained effects of disorder, photoexcitation, pressure[11] and misfit strain in high-temperature superconductors (and also pnictides[26]) can now be explored as potentially being due to granular fractal microstructure. The data also raise the more intriguing question of whether the oxygen defects order on fractal networks because the electrons that form the strange metal constitute a 'fractal glue', perhaps generated as a photographic-like image of the quantum critical charge fluctuations[27] sampled during annealing (at temperatures well within the non-Fermi-liquid regime). Another possibility is that fractal defect structure promotes superconductivity, by means of either conventional percolation or the more unconventional mechanism of Feshbach-like shape resonances[28] associated with anisotropic superconducting gaps in granular heterostructures[29].

**Acknowledgements** We are grateful to the ID13 beamline staff at ESRF, R. Davies, S. Agrestini, V. Palmisano, E. J. Sarria, L. Simonelli and A. Vittorini Orgeas for help in the early stage of this research project. We thank J. Zaanen and G. Bianconi for suggestions, comments and help with the data analysis. This experimental work has been carried out with the financial support of the European STREP project 517039 "Controlling Mesoscopic Phase Separation" (COMEPHS) (2005–2008) and Sapienza University of Rome, research project "Stripes and High-T c Superconductivity".

**References**

1. Dagotto, E. Complexity in strongly correlated electronic systems. Science 309, 257–262 (2005).
2. Tokura, Y. Critical features of colossal magnetoresistive manganites. Rep. Prog. Phys. 69, 797–851 (2006).
3. Wadhawan, V. K. Smart Structures: Blurring the Distinction Between the Living and the Nonliving (Monogr. Phys. Chem. Mater. 65, Oxford Univ. Press, 2007).
4. Bishop, A. R. HTC oxides: a collusion of spin, charge and lattice. J. Phys. Conf. Ser.108, 012027 (2008).
5. Müller, K. A. in Superconductivity in Complex Systems (eds Muller, K. A. & Bussmann-Holder, A.) 1–11 (Structure and Bonding 114, Springer, 2005).

6. Bednorz, J. G. & Muller, K. A. Possible high T c superconductivity in the $Ba_2La_2Cu_2O$ system. Z. Phys. B 64, 189–193 (1986).

7. Phillips, J. C. Percolative theories of strongly disordered ceramic high temperature superconductors. Proc. Natl Acad. Sci. USA 107, 1307–1310 (2010).

8. Skinner, S. J. & Kilner, J. A. Oxygen ion conductors. Mater. Today 6, 30–37 (2003).

9. Liu, Q. Q. et al. Enhancement of the superconducting critical temperature of $Sr_2CuO_{3+\delta}$ up to 95 K by ordering dopant atoms. Phys. Rev. B 74, 100506(R) (2006).

10. Frello, T. et al. Dynamics of oxygen ordering in $YBa_2Cu_3O_{61x}$ studied by neutron and high-energy synchrotron X-ray diffraction. Physica C 282–287, 1089–1090 (1997).

11. Lorenz, B., Li, Z. G., Honma, T. & Hor, P. H. An intrinsic tendency of electronic phase separation into two superconducting states in $La_{2-x}Sr_xCuO_{4+y}$. Phys. Rev. B 65, 144522 (2002).

12. Liu, L., Che, G., Zhao, J. & Zhao, Z. Thermal treatment effect of the oxidized $La_2CuO_4$: the access of continuous and discontinuous. Physica C 425, 37–43 (2005).

13. Mohottala, H. E. et al. Phase separation in superoxygenated $La_{2-x}Sr_xCuO_{4+y}$. Nature Mater. 5, 377–382 (2006).

14. Kusmartsev, F. et al. Transformation of strings into an inhomogeneous phase of stripes and itinerant carriers. Phys. Lett. A 275, 118–123 (2000).

15. Lee, Y. S. et al. Neutron scattering study of the effects of dopant disorder on the superconductivity and magnetic order in stage-4 $La_2CuO_{4+y}$. Phys. Rev. B 69, 020502(R) (2004).

16. Barabasi, A.-L. & Stanley, H. E. Fractal Concepts in Surface Growth (Cambridge Univ. Press, 1995).

17. Soh, Y.-A. et al. Local mapping of strain at grain boundaries in colossal magnetoresistive films using X-ray microdiffraction. J. Appl. Phys. 91, 7742–7744 (2002).

18. Evans, P. G., Isaacs, E. D., Aeppli, G., Cai, Z. & Lai, B. X-ray microdiffraction images of antiferromagnetic domain evolution in chromium. Science 295, 1042–1045 (2002).

19. Cojoc, D. et al. Scanning X-ray microdiffraction of optically manipulated liposomes. Appl. Phys. Lett. 91, 234107 (2007).

20. Daubechies, I. Ten Lectures on Wavelets (CBMS-NSF Regional Conf. Ser. Appl. Math. 61, Society for Industrial and Applied Mathematics, 1992).

21. Poccia, N., Ricci, A. & Bianconi, A. Misfit strain in superlattices controlling the electron-lattice interaction via microstrain in active layers. Adv. Condens. Matter Phys. 2010, 261849 (2010).

22. Agrestini, S., Saini, N. L., Bianconi, G. & Bianconi, A. The strain of $CuO_2$ lattice: the second variable for the phase diagram of cuprate perovskites. J. Phys. Math. Gen. 36, 9133–9142 (2003).

23. Hammel, P. C. et al. Localized holes in superconducting lanthanum cuprate. Phys. Rev. B 57, R712–R715 (1998).

24. Savici, A. T. et al. Muon spin relaxation studies of incommensurate magnetism and superconductivity in stage-4 $La_2CuO_{4.11}$ and $La_{1.88}Sr_{0.12}CuO_4$. Phys. Rev. B 66, 014524 (2002).

25. Clauset, A., Shalizi, C. R. & Newman, M. E. J. Power-law distributions in empirical data. SIAM Rev. 51, 661–703 (2009).

26. Caivano, R. et al. Feshbach resonance and mesoscopic phase separation near a quantum critical point in multiband FeAs-based superconductors. Superconduct. Sci. Technol. 22, 014004 (2009).

27. Cubrovic, M., Zaanen, J. & Schalm, K. String theory, quantum phase transitions, and the emergent Fermi liquid. Science 325, 439–444 (2009).

28. Bianconi, A. Feshbach shape resonance in multiband superconductivity in heterostructures. J. Superconduct. Novel Magnet. 18, 25–36 (2005).

29. Bianconi, A., Valletta, A., Perali, A. & Saini, N. L. Superconductivity of a striped phase at the atomic limit. Physica C 296, 269–280 (1998).

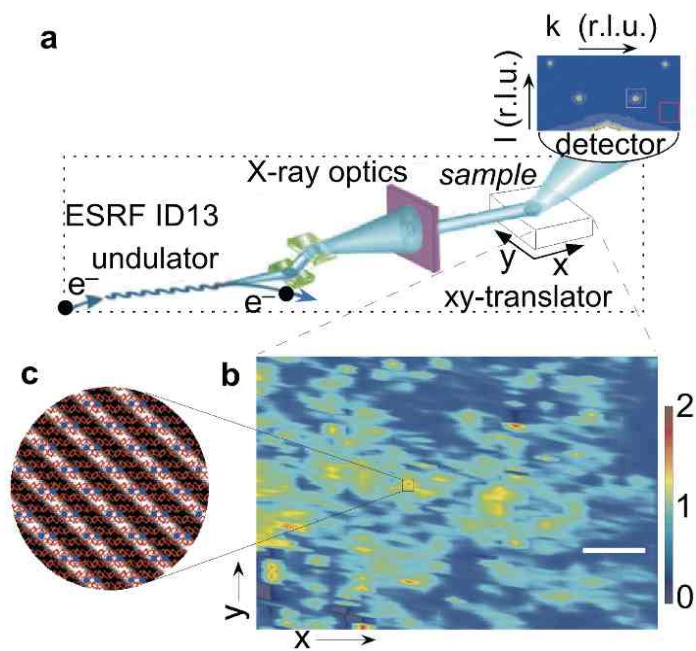

**Figure 1.** Mixed real- and reciprocal-space images of dopant ordering. **a**, The X-ray microdiffraction apparatus is located at the European Synchrotron Radiation Facility (ESRF) and features an electron undulator providing 12–13-keV X-rays to crystal optics followed by a tapered glass capillary, which produces a 1 μm$^2$ beam spot at the sample. A charge coupled area detector (CCD; right hand side) records the X-rays scattered by the sample. The intensity, $I(Q2)$, of the superstructure satellites due to the Q2 ordering of oxygen interstitials in the $La_2CuO_{4.1}$ crystal is integrated over square subareas of the images recorded by the CCD detector in reciprocal-lattice units (r.l.u.) and then normalized to the intensity ($I_0$) of the tail of the main crystalline reflections at each point (x, y) of the sample reached by the translator. **b**, Incommensurate order is highly inhomogeneous, even for an optimal ($T_c = 40$ K) superconducting sample of $La_2CuO_{4.1}$. The intensities of the superstructure satellites are presented on a logarithmic scale as a false-colour image. The scale bar corresponds to 100 μm. The intense red–yellow peaks in the two-dimensional colour map represent locations in the sample with high strength of the three-dimensional i-O ordering, and dark blue indicates spots of disordered i-O domains. The scanning images show few regions with intense satellite μXRD reflections and many regions with weak satellite μXRD reflections. **c**, Real-space view of the ordered domains that give rise to the Q2 superstructure imaged on the CCD detector. It highlights the i-O ions (blue dots) in the c–b plane of the Fmmm crystal structure of $La_2CuO_4$. The i-O located at the (1/4, 1/4, 1/4) site in the $La_2O_{2+y}$ spacer layers pair to form linear stripes in the orthorhombic a direction with a period of nearly four lattice units along the b axis in the a–b plane. The stripes alternate in different layers with a c-axis periodicity of two lattice units. The red octahedra indicate the $CuO_6$ octahedral coordination units in the $CuO_2$ plane.

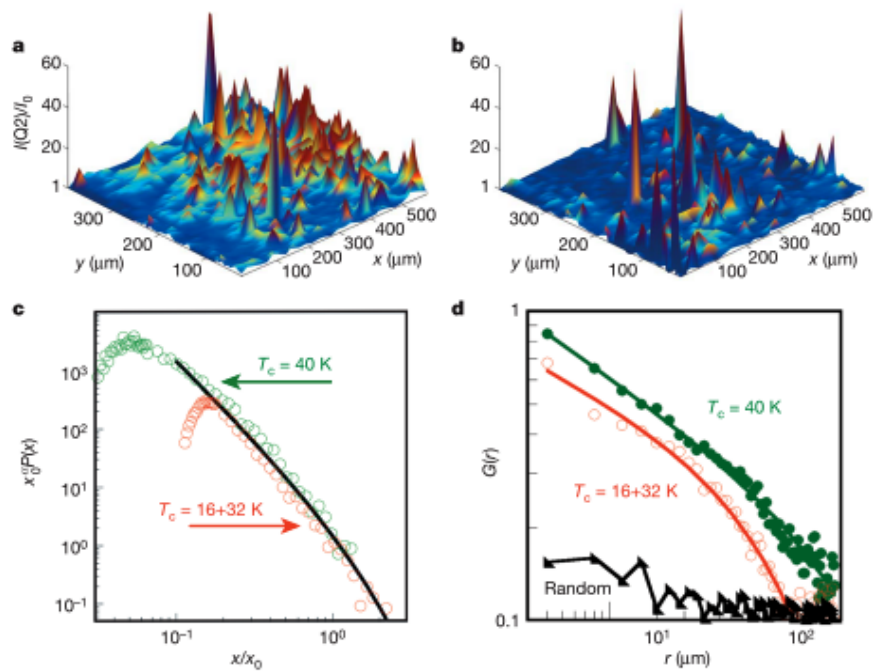

**Figure 2**. Scale-free fractal distribution and power-law statistical analysis of ordered i-O domains. **a, b**, The position dependence of the Q2 superstructure intensity $I(Q2)/I_0$ for two typical samples obtained by following different annealing–quenching protocols, resulting in $T_c = 40$ K (a) and $T_c$ 5 16+32 K (b) phases. Visual inspection of a and b shows that the spikes corresponding to ordered microdomains are more isolated for the more disordered sample with lower $T_c$ than for the high-$T_c$ sample, indicating that the nucleation and growth of Q2 regions proceeds to smaller length scales for shorter annealing times. **c**, The probability distribution, $P(x)$, of the Q2 XRD intensity $x = I(Q2)/I_0$ scales at sufficiently high intensity as a power-law distribution with exponential cut-off $x_0$. The data are fitted by the function described in the text. The fitted power-law exponent is given by $\alpha = 2.6 \pm 0.2$ independently of the sample critical temperature, and the cut-off increases from $7 < x_0 < 9$ for the $T_c = 16+32$ K samples to $28 < x_0 < 33$ for the $T_c = 40$ K samples. In the plot, we show that the $P(x)$ distributions, when properly rescaled, collapse on the same universal curve (solid line). **d**, Spatial correlation function, $G(r)$, where $r = \left|\vec{R}_i - \vec{R}_j\right|$, calculated for the intensities at the spots $\vec{R}_k$ mapped in a and b. The spatial correlation function does not have the standard exponential behaviour but instead obeys a power law, $G(r) \propto r^{-\eta} \exp(-r/\xi)$, with cut-off $\xi$, as expected, near a critical point (see text for details). The correlation length, $\xi$, increases with increasing $\langle I \rangle$, and in the illustrated cases for $T_c = 16+32$ K and 40 K has respective values $180 \pm 30$ μm and $400 \pm 30$ μm.

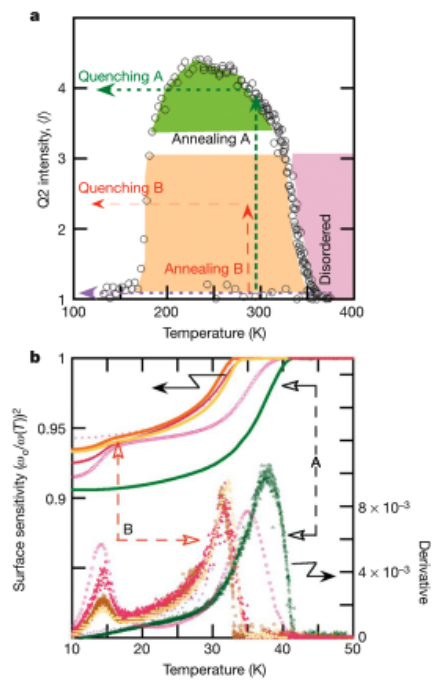

**Figure 3**. Nucleation and growth of i-O superstructures. **a,** The integrated intensity, <I>, of the Q2 superstructure satellite reflections as a function of temperature measured using a large (200 μm diameter) beam at the Elettra storage ring. Starting from 300 K, we warmed the sample at a slow heating rate, 10 K h 21. The decreasing <I> becomes indistinguishable from zero at 350 K, indicating a continuous order-to-disorder phase transition in the 310–350-K range. A rapid quench from the disordered phase to below 200 K produces a stable sample with randomly distributed i-O (dashed red arrow). Heating the sample very slowly induces a disorder-to-order transition in the 180–200-K range. Annealing the random i-O sample for a long time at constant temperature in the 200–300-K range (following the green vertical arrow) allows the i-O ordering to reach the maximum allowed intensity of the Q2 superstructure (we used annealing process A to produce the sample image in Fig. 2a). Using annealing process B in the 200–300-K range followed by quenching process B produces samples with reduced i-O domain order (we used annealing process B for the sample in Fig. 2b). **b,** Temperature-dependent complex conductivity of doped $La_2CuO_{4.1}$ superconducting crystals. We used a single-coil inductance method recording $(\omega_0/\omega(T))^2$, where ω(T) is the resonance frequency of an LC circuit, L is the inductance of the submillimetre coil placed near the surface of the superconducting sample and $\omega_0$ is the reference resonance frequency for a non-superconducting sample. The lower part of the figure shows the derivatives of $(\omega_0/\omega(T))^2$. The yellow, orange and red curves are for the two T c = 16+32 K samples, where $T_{c1}$ = 16 K, 32 <$T_{c2}$ < 36 K and <I> = <I(Q2)/$I_0$> is small (annealing process B), and the green curves are for the $T_c$ = 40 K samples, where <I> = <I(Q2)/$I_0$> is large (annealing process A).